\documentclass{aa}
\usepackage{psfig}

\newcommand{\norm}[1]{\mid #1 \mid}
\newcommand{\der}[2]{\frac{d#1}{d#2}}
\def\spt {space-time}
\def\calM {\cal{M}}
\def\calL {{\cal L}}
\def\calN {{\cal N}}
\def\calO {{\cal O}}
\def\calD {{\cal D}}
\def\calC {{\cal C}}
\def\calD {{\cal D}}
\def\Ret{{\rm I\!R}^{3}}

\def\RR {{\bf R}}

\def\eg {{e.g.}}
\def\ie {{i.e.}}

\def\dS{de~Sitter}

\def\EdS{Einstein -- de~Sitter}
\def\frl{Friedmann-Lema\^ \i tre}

\def\gr {general relativity}

\def\spt {spacetime}
\def\mink {Minkowski spacetime}

\def\hata {\hat{a}}

\def\hate {\hat{e}}

\title{Space and Observers in Cosmology}
\titlerunning{Space and Observers}
\begin{document}
%\begin{opening}
\author{M. ~Lachi\`eze-Rey\\
 Service d'Astrophysique, C.E. Saclay,\\
  F-91191 Gif sur Yvette Cedex, France} 
\authorrunning{Lachi\`eze-Rey} %\institute{Service d'Astrophysique, C. E. Saclay\\91191 Gif sur Yvette cedex, 
%France}}
%\end{opening}

%\pagestyle{empty}
%\tableofcontents
\maketitle
\abstract{I provide a prescription to define space, at a given moment, 
for an arbitrary observer in an arbitrary  (sufficiently regular) 
curved \spt. This prescription,  based on synchronicity (simultaneity) 
arguments, defines a  foliation of \spt, which 
corresponds to a family of canonically associated observers. It 
provides also a natural global reference frame (with space and time coordinates) 
for  the observer,  in \spt ~(or  rather in the part of it which is  causally 
connected to him), which remains Minkowskian along his world-line. This 
definition intends to provide a basis for the problem of 
quantization   in curved \spt, and/or for non inertial observers.

Application to \mink ~illustrates clearly the fact that different 
observers see different spaces. It allows, for instance, to define 
space everywhere without ambiguity, for the Langevin observer 
(involved in the Langevin pseudoparadox of twins). 
Applied to the Rindler observer (with uniform acceleration) it leads 
to the Rindler coordinates, whose choice is   so justified with a 
physical basis. This leads to an interpretation 
of the Unruh effect, as due to the observer dependence of the definition of 
space (and time).

This prescription  is also applied in cosmology, for inertial observers
in the  \frl ~models: space for the observer appears to differ from 
the hypersurfaces of homogeneity, which do not obey the simultaneity 
requirement. I work out   two examples: the \EdS ~model, in which 
space, for an inertial observer, is not flat nor homogeneous, and the 
\dS ~case.
}

\section{Introduction}
%%%%%%%%%%%%%%%%%%%%%%%%%%%%%

General relativity and relativistic cosmology consider {\sl \spt} as 
the  arena for physics and it is an old question to define   {\sl space} 
and {\sl time}.  These notions are   not  covariant and 
all problems of \gr ~and  cosmology can be addressed without them, so 
that they  may  appear  as rather academic (this 
is equivalent to the choice of a {\sl global reference frame}). 
However, on the one hand, the literature refers  often to space, for instance 
when  it is   affirmed that space is flat (or not), or 
homogeneous (or not) in a given 
cosmological model.  On the other hand,  quantum physics, or its 
interpretation,   requires most 
often a splitting of \spt ~into space and time. This points out the  necessity   
of a  convenient definition of   space in \spt, or quivalently, the 
choice of a convenient reference frame. The simple  example of two 
inertial  observers in   \mink, with different velocities, shows that such 
a definition  must be  observer-dependent. 

An observer needs  a frame to do physics in his environment. Although   space 
is defined  {\sl locally}  without  ambiguity (by  orthogonality to his 
worldline, \ie, to his velocity $u$),
there are many different ways to extend this definition, \ie, to choose  a frame
beyond a neighborhood. Since many experiments (\eg, synchronisation procedures)  
involve the observer at  different  moments of his history, a   global frame is   
constrained to have convenient properties 
along his world line $O$, or  at least  a part of it (like being 
Minkowskian there). But, again, this is far 
from  being   sufficient to determine the choice. 
This paper provides a prescription which  associates to any  observer, defined by his 
world line, a unique   reference frame with special properties.

A reference frame involves, as a first step,  a family of  spatial hypersurfaces 
orthogonal to $O$, which can be chosen in many different ways. Among 
popular prescriptions, the  
{\sl Fermi coordinates} require  the 
hypersurfaces  to be  generated by    the spacelike  geodesics orthogonal  
to $O$. But, as  it is well known,   these spatial 
surfaces  may intersect, even in simple situations like for the Langevin 
observer (see below). This   forbids a definition of time valid far from  $O$ 
(different   values of time would be  associated to the same event).
In cosmology, another   popular  choice   is to select   the 
orthogonal spatial hypersurfaces  of homogeneity. But such 
hypersurfaces do not exist in all \spt s, and the choice is clearly 
not convenient when the observer himself breaks the spatial 
symmetries (by his acceleration or rotation  for instance).
The prescription  proposed here  does not suffer from these drawbacks.  In 
addition, it is the only one to obey the  
"~simultaneity criterion~",   absent in all other prescriptions (excepted 
in the  immediate  neighborhood of $O$) :    the different points of "space"   
are   seen  as simultaneous (see below for a pecise definition) by the observer. 
Moreover, it has the 
additional  advantage to depend only on the conformal structure of the 
metric (except for the proper time of the observer), and its validity is  broader than 
those mentioned above (for instance, in the absence  of spatial homogeneity).

This prescription has in fact already been applied {\sl de facto} in some 
circumstances, like the Rindler coordinates.  But its validity is much 
more general, without too  restrictive conditions: for any observer 
(inertial or not), in any \spt ~(with 
some conditions, see below), it defines uniquely a "~simultaneity 
space~" (shortly, space) $\Sigma _{\tau}$ at any moment $\tau$.  The 
$\Sigma _{\tau}$ do not intersect (even in the situations where the Fermi 
hypersurfaces do, like for instance for the Langevin observer, see 
below), and they are defined even in the absence of spatial homogneity.  
This allows to extend the validity of the observer's proper time to 
the  
whole \spt.  In the \frl ~models, space so defined does not coincide 
with the intuitive idea of what space could be.

%~(assumed to be sufficiently regular, see below).   

Numerous attempts to  define a quantization procedure in curved  \spt, 
and/or  for non inertial  observers   (see, \eg, Birrel   and 
Davis, 1982), % \cite{Birrel}), 
involve, more or less explicitly, 
a space-time splitting of \spt. This is especially  important  for 
giving a physical interpretation of   quantum states in terms of  
frequencies or    particles.  For instance,  I   show below   that the
current interpretation of the Unruh effect, based on the Rindler coordinates, 
correspond in fact to the prescription 
promoted here.  

I emphasize that  all quantities introduced in this work are 
covariant. This includes all the observer dependent quantities like 
his   velocity, acceleration,   world-line and the special  reference frame 
introduced here.  An  observable quantity (\eg, the energy)
  is a combination   of a covariant  quantity  associated to the 
observer  (\eg, its velocity $u$) with a  
covariant quantity  associated to the observed  system 
(\eg, its  momentum-energy tensor). The definition of a global frame 
associated to an   observer may allow 
to define properly non local quantities related to him.

Since the  goal of this work is to construct a reference frame for an observer in 
cosmological situation, one only requires its validity in the part 
of \spt ~which is causally related to him:  all events considered 
here are  causally linked to the observer. 
Throughout this paper, by an abuse of langage,  I will call "~\spt ~" 
the set $\calM  _{0}$  of events 
inside the  particle horizon and   the event horizon of the observer, if 
they exist.  This first restriction 
applies to  any  construction of a  global  reference frame. Second, the synchronicity 
arguments require that   null geodesics 
admit no conjugate points. This excludes the presence  of gravitational 
lensing (also incompatible with the other prescriptions). 
Since this   analysis concerns an observer in a  cosmological 
situation (convenient for  quantization), or for  studying the 
perturbative  development  of  irregularities, this condition is  not 
too   restrictive  (I emphasize again, that the validity of this 
approach is broader than for concurrent proposals). 

In Section 2, I will implement the definition of simultaneity space, and 
the related notions. I show how they allow to define  a global frame 
convenient  to the 
observer, and a   congruence of  associated observers. 
Section 3  applies these results to the inertial and Langevin 
observers in \mink. Section 4 considers the Rindler observer in 
\mink, and the associated Unruh effect. Section 5  considers the 
\frl ~cosmological models.

\section{A global reference frame for observers}
%%%%%%%%%%%%%%%%%%%%%%%%%%%%%

\subsection{The accelerated Observer}
%%%%%%%%%%%%%%%%%%%%%%%%%%%%%

The most general  observer is defined as a timelike world line $O(\tau)$ 
parametrized by  proper time $\tau$. This defines its velocity $u (\tau)$ and its 
acceleration $a (\tau):=\nabla _{\tau}u$ everywhere on his world-line.
%We will consider special cases of observers (geodesic, Killing, 
%conformal Killing,\ldots) and special cases of the metric 
%(static,\ldots) but here we consider the most general observer. 
I call $\hata$ the unit vector parallel (and in the same 
direction) to the  acceleration: \begin{equation}
a:=\der{u}{\tau}=\hata~\norm{a}. \end{equation}
Herereafter I consider only {\sl non rotating} observers for which  
$\nabla _{\tau} \hata:= \nabla _{\tau} \hate_{1}
=\norm{a} ~u $ (all the procedure here applies to the case with 
rotation, which  will be considered in a 
subsequent paper). Transport along the world line corresponds to a Lorentz rotation 
(in the plane  $u,a$).

Defining $\hate _{0} (\tau)=u (\tau)$ and $\hate _{1} (\tau)=\hata 
(\tau)$, the transport is expressed by    \begin{equation}
\nabla _{\tau} \hate _{A}=  \Omega.\hate _{A}~~~~(A=1,2),
\end{equation} where  $(\Omega .V)^{\mu}:= a^{[\mu}~u^{\nu]}~V_{\nu} $ for the observer 
without (spatial) rotation. 

We can associate naturally two different  orthonormal  frames  to the 
non rotating   
observer, along his world line. 
First, there is  a parallely transported frame $E$ but,   for a non 
inertial observer,  no vector of this frame coincides with the velocity. 
Second, there is a   Fermi-transported frame $f$ such that $f_{0}:=u$. 
Since there is no spatial rotation, $f _{1}=a/\norm{a}$ and the law 
$\nabla _{\tau} \hate _{A}=  \Omega.\hate _{A}$ is verified  for 
$A=0,1,2,3$. 

For  {\sl non rotating observers}, the spatial sections 
(or spaces)  keep axial symmetry around the unique spatial  direction defined by 
the acceleration  vector.  
%This makes the problem  essentially
%two  dimensional, in the sheet defined by $u$ and $a$ (the rest of 
%\spt ~being generated by axial symmetry).  
For inertial observers (mainly considered in this paper, except  in  \ref{Langevin} 
and \ref{Rindler}),  the  problem is  purely  2-dimensional, with spatial spherical symmetry. 
The general (4-dimensional)  case, including  the rotating observer will be 
examined  in a subsequent paper (most of the   calculations  presented here will
 remain  valid).

\subsection{Definition of space}
%%%%%%%%%%%%%%%%%%%%%%%%%%%%%%%%

A given \spt ~admits, in general,      many  time-like foliations   
with associated   space-like sections,   compatible
 with the  world-line of a 
given observer $O$. I will show that the requirement of 
synchronicity allows to select a unique one, and thus provides a 
special global  definition of   {\sl  space}.  This 
will require to consider the  whole story of the observer, including his  future. 
In the following, I  will assume that  the   causal structure of \spt ~ 
admits   only  
light cones without foldings and  conjugate points.  
%(this excludes 
%a metric giving rise to   gravitational lensing). 
These  restrictions are  however  
appropriate for cosmology, and characterize  a background \spt ~convenient for 
quantization. This excludes also multi-connected ~\spt s. In the 
whole paper, I  denote $\tilde{v}$ the one-form metric-dual to a 
vector $v$, \ie, such that $<\tilde{v},v>=g(v,v):=\norm{v}^{2} $.

It is well known that it is impossible to define absolute simultaneity 
in special or \gr. However, a {\sl local}  prescription of  simultaneity or, 
better, {\sl synchronicity}, {\sl from the point of view of an 
observer}   is 
widely  used (see,  \eg, Landau and  Lifshitz, 1966),
%\cite{Landau}), 
and   defines a {\sl  local} space-time splitting  for this observer. The 
construction of space in this work is based to the extension of this 
prescription beyond a local neighborhood, with arguments of synchronicity 
which have a perfectly  operational character. 

Given an  observer $O$, I define $\Sigma _{ \tau}$, the {\it hypersurface of 
synchronicity} (HS) of $O$ at  proper time $\tau$, as the set of events 
related by a null geodesics to  both $O(\tau+\delta)$ and  $O(\tau-\delta)$, 
where $\delta$  is an arbitrary interval of proper time for $O$: \begin{equation}
\Sigma _{ \tau}=\cup _{\delta} ~[I^{future} (\tau-\delta)    
\cap I^{past} (\tau+\delta)  ],\end{equation} where $I^{past} 
(\tau)$ [resp. $I^{future} (\tau)$] denotes the null past [future] 
light-cone of the observer at proper time $\tau$.

\begin{figure} 
\psfig{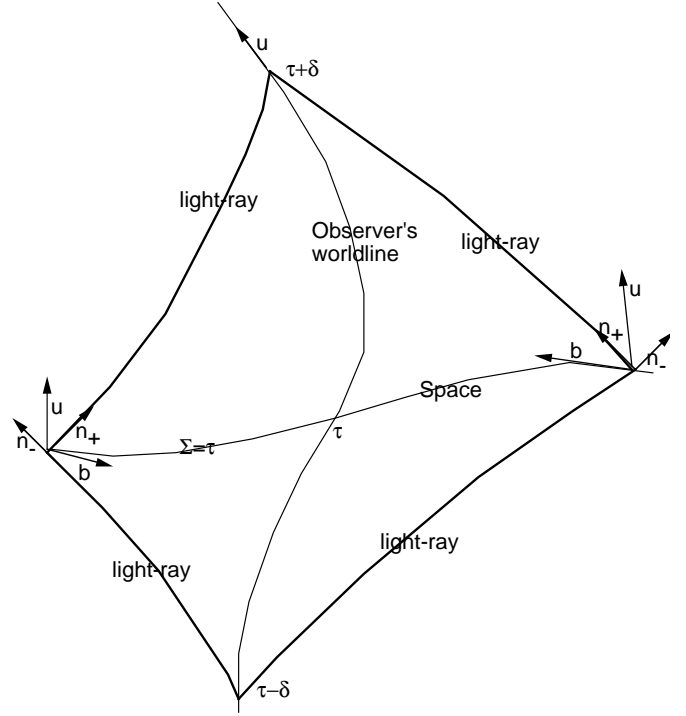}
\caption{At each point $x$, the light-rays from the observer (in the past 
and in the future) define the vectors $n^{+ }$ and $n^{-}$. The 
velocity $u$ of the observer is (non parallely) transported to give 
the vectors $u$ and $b$ at $x$.}
\label{Fig1}
%\label{EdSSection}
   \end{figure}

Given the restrictions above, the surfaces $\Sigma _{ \tau}$ for different values 
of $\tau$  completely fill     $  \calM  _{0}  $. This allows to extend 
the vector field $u$ to the totality of $\calM _{0}$ by requiring that 
it is everywhere unit ($u.u=1$) and orthogonal to $\Sigma (\tau)$ 
(Fig.\ref{Fig1}). The 
vector field $u$   constitutes a  foliation of $\calM _{0}$, with 
the $\Sigma _{ \tau}$   as transverse (orthogonal) surfaces. Each integral line 
of  $u$ will be labelled by its intersection $\RR$ with $\Sigma (1)$ 
(for instance), so 
that any point of   $\calM _{0}$ can be written $(\tau,\RR)$. 
Practically, $R:= \norm{\RR}$ (evaluated in the spacelike surface  $\Sigma _{ 
\tau}$, see below) can be seen as half the interval of  proper time  
between the  emission of a flash light which illuminates a cosmic object 
at $x$, and the 
observation of the resulting image; or, equivalently, as half the interval of  
proper time  between the 
emission of a light ray (or radar signal)  by an observer, and its reception after 
mirror reflection by the object at $x$. The requirement
 of synchronicity imposes that  " the space for $O(\tau)$ " is  $\Sigma 
_{ \tau}$.

Smarr and York (1978), among others, have considered space - time 
slicing of \spt ~associated to a congruence of observers, what they 
call a kinematics of \spt.  The foliation introduced here is of this 
kind, associated with the congruence of " associated observers ". 
However, since the latter are defined from the world line of the 
unique observer $O$, this procedure defines an unique space - time 
slicing,  from the  world line  of  this unique observer only.  In a 
forhtcoming paper, the properties of this " kinematics " will be 
discussed in more detail, and compared with other possibilities. 

%%%%%%%%%%%%%%%%%%%%
%%%%%%%%%%%%%%%%%%%%
%%%%%%%%%%%%%%%%%%%%
%%%%%%%%%%%%%%%%%%%%
%%%%%%%%%%%%%%%%%%%%
%%%%%%%%%%%%%%%%%%%%

 The integral lines of $u$ define an unique family of observers 
 associated to $O$, that I will call {\sl the canonical observers} 
 associated to  $O$. Care must be taken that they do necessarily  not share 
the properties of $O$. For instance, they are not necessarily 
geodesic, even when  $O$ is,  when there is expansion. Along 
the word lines generated by $u$ we will call the   proper time $t$, 
with %so that $u:=\nabla _{t}$ everywhere. 
  $t=\tau$ along $O$. 

\subsection{Transport along the light-rays}
%%%%%%%%%%%%%%%%%%%%%%%%%%%%%%%%{\bf }

The properties of space are defined from those of the world-line of 
the observers, transported by the past and future light-rays. To 
explore them, we define the  two  null functions $\calN _{-}(x)$ and 
$\calN _{+}(x)$ such that 
the value of $\calN _{-}(x)$ [resp. $\calN _{+}(x)$] 
is the    proper time $\tau-\delta$ of the observer $O$, when it emits a 
light-ray reaching $x$  [resp. $\tau+\delta$,  when it receives  a 
light-ray emitted from $x$]. In other words the null hypersurface     
$\calN _{-}(x)=\tau$  [resp. $\calN _{+}(x)=\tau$] 
is the future [resp. past]   light cone  of the observer at proper time 
$\tau$.     We define   their (null)   
(past and future) generators as   $\tilde{n} _{\pm } =\nabla  \calN _{ 
\pm}=d  \calN _{ \pm}$. Both are 
future directed, and normalized so that the frequency emitted   or 
received by the observer is unity (see below).

It is easy to show that $\Sigma _{\tau}$ is defined by the  equation 
\begin{equation} T(x):= [\calN 
_{-}(x)+\calN _{+}(x)]/2=\tau.\end{equation}  
For any point, $T$ constitutes a natural time-coordinate.
In addition we   define the deformed cylindric  
hypersurface \begin{equation} 
R(x):= [  \calN _{+}(x)-\calN _{-}(x)    ]/2=\delta
\end{equation} as the set  of events at a constant  "~proper time  
interval~" 
$\delta$ from the observer, when he  describes its world-line. We define the 
proper time interval (PT interval) $\delta$ of an 
event $x$ as half the  observer's  proper time interval  between the instants 
$\tau _{1}=\tau +\delta$ and $\tau _{2}=\tau -\delta$, when he receives and emits 
the light  rays emitted and received by the event $x$. For any 
point,   $R=\norm{\RR}$ defines a  natural radial space coordinate.

 Given the normalization 
above, we have \begin{equation}
d T=(\tilde{n}_{-}+\tilde{n}_{+})/2  ~\mbox{and } 
d R=  (\tilde{n}_{+}-\tilde{n}_{-})/2. \end{equation} It 
is easy to check that $d T \cdot d R =0$, and \begin{equation}
d T \cdot d T=-d R \cdot d R:=N^{-2}= n_{+} \cdot  n_{-}/2,
\end{equation} which defines the {\sl lapse function} $N$ associated 
to this foliation. Since $dT$ is  orthogonal to $\Sigma $,
 we have $\tilde{u}=N ~d T$.    From $u^{2}=1$, we 
have $N~u \cdot  d T =1$. Since, along $\calO$, $ \tau=T$, this implies $N=1$ on $\calO$.
%As we will see, $N(x)$  is the  red or blue shift attached to the 
%light received by thge observer from $x$. It is also the shift of the 
%light received by $x$ from the observer.

Everywhere (except on $\calO$),  we define \begin{equation}
\tilde{b}:=N ~d R =\tilde{u}
- N~\tilde{n}_{-} =-\tilde{u}+ N~\tilde{n}_{+}.\end{equation}
 We have    $ b^{2}=-1$,  $u\cdot b=0$, $u+b= N~n_{+}$ and $u-b= N~n_{-}$. 
 Thus, $b$ is a unit
space like vector, tangent to $\Sigma _{\tau}$ and, in some sense, 
pointing towards the observer $O$.  In general, the vector $b$ is not 
geodesic but it is {\sl chorodesic}, due to the 
synchronicity property and the congruence of associated observers is 
{\sl quasi-rigid}  (see Bel, 1998).%\cite{Bel}). 

\subsection{Associated observers}
%%%%%%%%%%%%%%%%%%%%%%%%%%%%%%%%

The vector field $u$ is perfectly defined everywhere and characterizes 
the family of canonical observers. This family  defines a " 
kinematics " in the sense of Smarr and York (1978). All the relevant  
formalism of projectors, lapse and shift functions (the latter being 
zero here, see below), intrinsic 
curvature, etc., applies and will be discussed in a forthcoming 
paper.

\begin{figure} 
\psfig{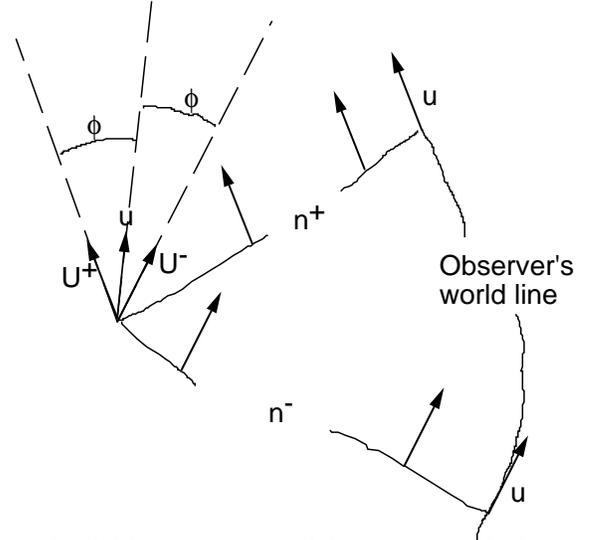}
\caption{The field $u$ is not parallely transported. At each  point 
$x$, $U^{+}$ is defined by the parallel transport of $u$ from the 
future, \ie, by  $n^{+}$,  $U^{-}$ is defined by the parallel transport of $u$ 
from the  past, \ie, by  $n^{-}$.  }
\label{Fig2}
%\label{EdSSection}
   \end{figure}

The vector fields $u$ and $b$ are not transported parallely along 
the light rays ($u$  is not parallely transported along itself, 
in general). To have a clearer view, it is 
convenient to introduce the two   vector fields $U^{+}$ and 
$U^{-}$ which are, by definition, parallely transported along $n^{+}$ 
and $n^{-}$ respectively, $ n^{\varepsilon}.\nabla U^{\varepsilon}=0$, and which
both  coincide with $u$ along the world line of 
$O$ (Fig.\ref{Fig2}). These requirements are sufficient to define them 
everywhere. Both are (time-like) unitary and, when $O$ is inertial, 
the spherical symmetry allows to    developed them     as \begin{equation}
 U^{\varepsilon}=\cosh \phi  ~u+\varepsilon ~\sinh \phi  ~b.
 \end{equation}
 
In the same way that we introduced the unit vector field $b$ orthogonal 
to $u$ and in the plane defined by the light-rays, we can introduce 
two unit vector fields $B^{+}$ and $B^{- }$, orthogonal to $U^{+}$ and 
$U^{-}$, and in the same plane. 
What is interesting here is that the values of $U^{+}$ and $B^{+}$, at 
any point $x$, depend  only (via the transport along the light ray $n^{+}$) on the
velocity of  the observer $O$ at the value  $T(x)+R(x)$ of its proper time.
For this reason, we could call  $(U^{+}, B^{+})$ the " future frame ". 
Similarly,  $U^{-}(x) $ and $B^{-}(x)$, the "~past frame~",  depend only on  the velocity of 
 $O$ at the value  $T(x)-R(x)$ of its proper time.  
The two frames   at point $x$ are   transformed into   each other by a Lorentz 
rotation of  (hyperbolic) angle $2\phi$ and the frame   $[u(x),b(x)]$ is a 
" bisector "   frame, obtained from the previous ones   by  hyperbolic rotations 
of angles $  \phi $ and  $  -\phi $ in the 
$(n^{+} ,n^{-} )$  plane: the value   $u(x)$ can be reconstructed as  \begin{equation}
u=\frac{U^{+}+U^{-}}{\sqrt{2~(1+U^{+} \cdot U^{-})}}.\end{equation}  

The geodesic equation $n^{+} \cdot \nabla 
n^{+} =0$ leads to $n^{+} \cdot \nabla (\ln N^{-1})
=  ~n^{+}.\nabla \phi  $ and, after  integration along the light-ray,  
\begin{equation}
\phi   (x)=-   \ln  N(x).\end{equation} Finally, $U^{\varepsilon}=[n 
^{\varepsilon}+N^{2}~n^{(-\varepsilon)}]/2$.
Note that 
$U^{\varepsilon}$ may be used like $u$ to construct spacelike 
hypersurfaces and reference frames. For instance, Marzlin (1994) %\cite{Marzlin} 
proposed a reference frame based on $U^{+}$, but which does not obey 
 synchronicity.

%%%%%%%%%%%%%%%%%%%%
%%%%%%%%%%%%%%%%%%%%
%%%%%%%%%%%%%%%%%%%%
%%%%%%%%%%%%%%%%%%%%
%%%%%%%%%%%%%%%%%%%%
%%%%%%%%%%%%%%%%%%%%

This example of a bisector frame, in relation to two other frames, 
may   be extended to provide  a   {\sl local}     surface of  
synchronicity   for two different observers (see Fig.5 of 
Ali et al., 1990), %\cite{Gazeau}),  
useful for the study of  the quantum evolution of two 
interacting  particles  in \spt ~(Ali et al., 1990). %\cite{Gazeau}). 
This   will be further 
developed in a subsequent paper.

There is a closed path starting from the arbitrary point $x$, going 
along the light-ray $n^{+}$ to the (inertial) observer, then going back to the 
past of the observer along his world-line, and back to the point $x$ 
along the light-ray $n^{-}$.
When the observer is inertial, $u$ is parallely transported along his 
world-line, so that  the transport of $U^{+}$, then $u$ and then  
of $U^{-}$ remains parallel. This angular deficit of   $2\phi$, along 
a closed path in   the 
sheet $\calD$  (see below) is a mark of the average curvature of \spt 
~there.

{\bf Redshifts}

Let us consider a congruence of (non necessarily canonical) objects  
with velocity $V(x)$ at the   point (event)  $x$  in \spt.  An  object  
at $x$ is 
seen with redshift $z^{+}=(n^{+}(x) \cdot V(x))^{-1} $ by the observer (in the future) 
and  sees  the observer (in his past) with a redshift $z^{-}=n^{-}(x) \cdot V(x)$.
For the congruence of canonical observers, $z^{+}=N (x)$ and $z^{-}=N(x)^{-1}$. 
These observers are comoving with respect to the coordinate $R$, \ie, 
they keep a constant value of $R$.
On the other hand, there is a unique congruence of   objects for 
which $z^{+}=1$ [resp. $z^{-}=1$], those with velocity $U^{+}$ 
[resp. $U^{-}$].

{\bf Frames}

Since we assume that the observer is non rotating, the problem has 
a spatial axial symmetry around the direction of his acceleration    
(and even spherical symmetry if the observer is inertial, 
\ie, without acceleration). 

\begin{figure} 
\psfig{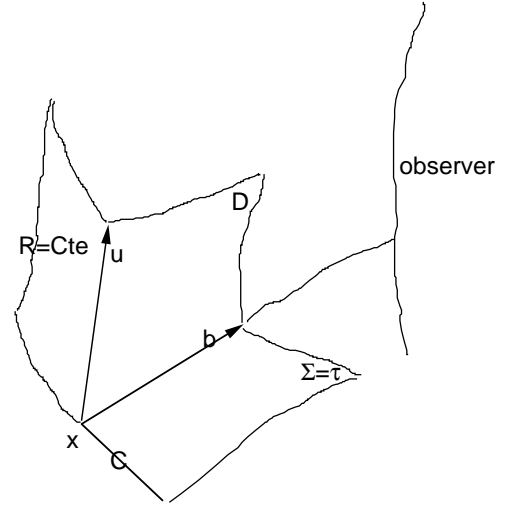}
\caption{At each  point 
$x$, $u$ and $b$ are tangent to the surface $\calD$. The curve $C$ 
(tangent to $b$ is contained in space $\Sigma _{\tau}$ and in the "~cylinder~" 
$R=C^{te}$.}\label{Fig3}
   \end{figure}

At each point $x \in \calM$ (except on $O$), the two vectors  
$n_{+}(x)$ and $n_{-}(x)$, or $u(x)$ and $b(x)$, or 
$U^{+}(x)$ and $U^{-}(x)$ define the same plane $D_{x} $ in the tangent space 
$T_{x} $. These  
planes  generate a foliation with  integral surfaces  $\calD$ 
(Fig.\ref{Fig3}). Each 
surface $\calD$  is  generated, with time, by a constant spatial direction 
$(\alpha,\beta)$ from the point of view of the  observer. In each 
$\calD$ (itself parametrized by $(\alpha,\beta)$),    
we will use $R$ as a spatial  (radial) coordinate and $T$ as a temporal 
coordinate. Thus any point   $x =(T,R, \alpha,\beta)$. 
 By definition, a light  ray from [to] a point 
$(T,R,\alpha,\beta)$ reaches [resp. starts from] the observer at proper 
time $T+R$ [resp.  $T-R$]. The coframe  $(dT,dR)$ has for dual frame   
$(e _{T}:=\partial _{T},
e  _{R}:=\partial _{R})$. These two vectors  generate the surface 
$\calD$. From  \begin{eqnarray}
<dT,e_{T}>=<dR,e_{R}>=1,\\ ~<dR,e_{T}>=<dT,e_{R}>=0,  
\nonumber \end{eqnarray} we can show  $e_{T}=N^{2}~(n_{+}+n_{-})/2=N~u$ and 
$e_{R}=N^{2}~(n_{+}-n_{-})/2=-N~b$. The two vectors  commute and thus they do 
Lie-transport each other: $e _{T} \cdot \nabla ~ e  _{R} = e  _{R} \cdot \nabla ~ e _{T}$.  
This is the same for the two vectors $N^{2}~n^{+}$ 
and $N^{2}~n^{-}$.

%%%%%%%%%%%%%%%%%%%%
%%%%%%%%%%%%%%%%%%%%
%%%%%%%%%%%%%%%%%%%%
%%%%%%%%%%%%%%%%%%%%
%%%%%%%%%%%%%%%%%%%%
%%%%%%%%%%%%%%%%%%%%

The    space-like plane $C_{x} $ orthogonal to all these vectors   is
tangent  to  the  past and future light cones at this point, and also to the 
hypersurface $\Sigma$ at this point. It is the tangent plane to the 
surface \begin{equation}
\calC=I^{+} \cap I^{-}=  I^{+} \cap \Sigma. \end{equation}
This    deformed 2-sphere  $\calC (\tau,\delta) $ is the set of all the 
points at  equal  PT-interval $\delta$ from the observer at the 
moment $\tau$, \ie, the set of all points of  the   hypersurface $\Sigma _{\tau}$   
at equal  PT-interval. Both $u$ and $b$ are  orthogonal to $\calC$.

We will choose two  orthogonal 
unit  space-like vectors $e_{3} $ and $e_{4} $ in $C$ (many choices 
are possible; since  we only consider  non rotating observers, the 
problem is essentially two-dimensional and we do not care about the  
choice of $e_{3} $ and $e_{4} $; this will be the subject of a forthcoming 
paper).    $e_{3} $ 
and $e_{4} $ form a basis for $C$, tangent to  the 2-surface $\calC$.
Any of the previous 
pair with these two vectors form a (pseudo-ON or ON) basis of the 
tangent space to \spt.  Also,   $n^{+}$ or $n^{-}$, $e_{3} $ and 
$e_{4} $  form   a pseudo-ON basis for the tangent space to the light 
cones.

{\bf The metric}

Since $u$ and $b$ are orthogonal and unitary, the metric can be 
written \begin{eqnarray}
ds^{2}=\tilde{u} ^{2}-\tilde{b} ^{2}-(e^{3})^{2}-(e^{4})^{2}\\
=N^{2}~(dT^{2}-dR ^{2})-(e^{3})^{2}-(e^{4})^{2},
\nonumber \end{eqnarray} where $-(e^{3})^{2}=g_{\alpha \alpha}~d\alpha^{2}$ and 
$-(e^{4})^{2}=g_{\beta \beta}~d\beta^{2}$. Thus, $N$ appears as 
the {\sl lapse function} associated to the foliation, or to the 
congruence of associated observers (the shift  vector being zero).
Since $N=1$ at the position of $O$, the metric is locally Minkowskian 
for him, everywhere on his worldline. The usual ADM formalism allows 
to define time and space projectors, as well as the  fundamental 
forms (metric and extrinsic curvature) on the surfaces $\Sigma 
_{\tau}$ (see, \eg, Smarr and York, 1978). 

As we have said, $T$, $R$ (with appropriate angular coordinates)  constitute a 
well behaved system of  coordinates and provide a global reference 
frame in $\calM _{0}$. Written in  these coordinates the coefficients 
of the metric reduce to $\pm N^{2}$. But $N=1$ on $O$ and is 
stationary (by spherical symmetry arguments) at $O$ if the observer 
is inertial. Thus, {\sl when the  observer is inertial}, $T$ and $R$ 
are {\sl normal 
coordinates} based on any point of his world line.

We point out that $T$ and $R$  are {\sl not}, in general,  {\sl Gaussian normal}   
coordinates since $T$ is not a proper time,
except on $O$ (Gaussian normal   coordinates cannot, in general, 
be constructed for this foliation). But they have the advantage that the 
surfaces 
of  constant "~time~" $T$ are global surfaces of simultaneity, although 
the surfaces of constant Gaussian normal   
coordinates  are not (they verify  only {\sl  local}   simultaneity).

Now I will consider some special cases.

\section{The inertial  observer in \mink}
%%%%%%%%%%%%%%%%%%%%%%%%%%%%%

The inertial  observer $O$ has zero acceleration and there is no cosmic 
expansion. Its velocity is defined 
by $$u^{0}=c ,~u^{1}=s  ,~u^{2}= u^{3}= 0,$$ where 
 $c:=\cosh\psi$ and $s:=\sinh \psi$, the {\sl rapidity} $\psi$ being a constant.
Its   world line  is
$x ^{0}=c~\tau,~x^{1}=s~\tau,~x^{2}= x^{3}= 0$.
A  light-ray passing  through $x$ reaches $O$  at proper time $\tau$ such that 
\begin{equation}
(x^{0}-c~\tau)^{2}=(x^{1}-s~\tau)^{2}+ (x^{2})^{2} + (x^{3})^{2}. 
\end{equation} For the arbitrary point $x$, it is convenient to define 
$x^{0}=\Delta ~\cosh \beta,x^{1}=\Delta ~ \sinh \beta$, so that
%$ \frac{x^{0}- x^{1}}  {c-  s}$ and $ \frac{x^{0}+  x^{1}}  {c+ s}$.
the solutions of this equation  give
 \begin{eqnarray}
&\calN  _{\varepsilon}=\\\Delta ~\cosh (\psi-\beta) +\varepsilon ~
& \sqrt{\Delta ^{2} ~\sinh (\psi-\beta) ^{2}+(x^{2})^{2} + (x^{3})^{2}}.
\nonumber \end{eqnarray}
%$$ max[ \frac{x^{0}- x^{1}}  {c-  s}, \frac{x^{0}+  x^{1}}  
%{c+ s} ],$$
%$$\calN  _{f}=min[  \frac{x^{0}- x^{1}}  {c-  s}, \frac{x^{0}+  x^{1}}  
%{c+ s} ] ,$$
\begin{equation}
 T =\Delta ~\cosh (\psi-\beta)=c~  x^{0} -s~x^{1}, 
\end{equation}
\begin{eqnarray}
R= \sqrt{\Delta ^{2} ~\sinh (\psi-\beta) ^{2}+(x^{2})^{2} + 
(x^{3})^{2}}\\=
\sqrt{( s~  x^{0} -c~x^{1} )^{2}+ (x^{2})^{2} + 
(x^{3})^{2}},\nonumber \end{eqnarray}
and $N=1$.

%%%%%%%%%%%%%%%%%%%%
%%%%%%%%%%%%%%%%%%%%
%%%%%%%%%%%%%%%%%%%%
\begin{figure} 
\psfig{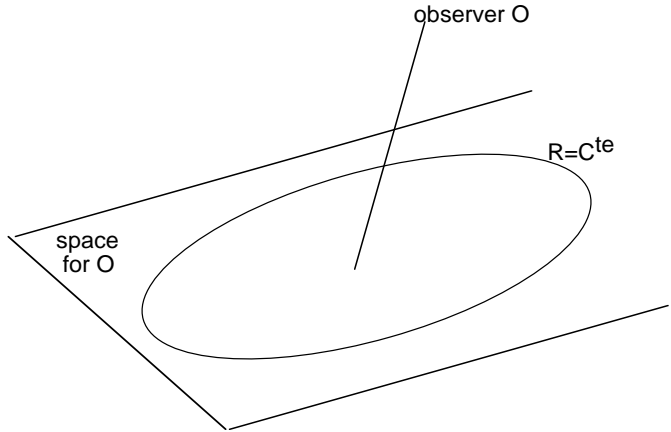}
\caption{For the inertial observer in \mink, 
with arbitrary velocity (rapidity $\psi$), space is the hyperplane 
with inclination $\psi$. We have drawn a curve $R=C^{te}$, in this 
plane.}
\label{Fig4}
   \end{figure}
%%%%%%%%%%%%%%%%%%%%
%%%%%%%%%%%%%%%%%%%%
%%%%%%%%%%%%%%%%%%%%

 The surface $\Sigma _{\tau}$ is the plane of  equation 
$c~  x^{0} -s~x^{1}~= \tau$, inclined by $\psi$ with respect to the 
vertical, and thus orthogonal to the world line of the inertial 
observer.   Thus, space is different for all inertial observers.

Finally,   $u= c~ d x^{0} -s~dx^{1}$ and 
 $$\tilde{b}=\frac{1}{R}~[( s~  x^{0} -c~x^{1} )~( s~  dx^{0} -c~dx^{1} ) + 
 x^{2}~dx^{2}+  x^{3}~dx^{3} ]  $$ points towards the observer at time 
 $\tau=T(x)$. In  the space $\Sigma _{\tau}$ (Fig.\ref{Fig4}), it is natural to define 
 the coordinate $y:=-s~  x^{0} +c~x^{1} =y/c$ so that $R^{2}=y^{2}+ 
 (x^{2})^{2} + (x^{3})^{2}$ and the spatial metric $-ds^{2}=dy^{2}+ (dx^{2})^{2}+(dx^{3})^{2} $.

\subsection{The Langevin  observer in \mink}\label{Langevin}
%%%%%%%%%%%%%%%%%%%%%%%%%%%%%

\begin{figure} 
\psfig{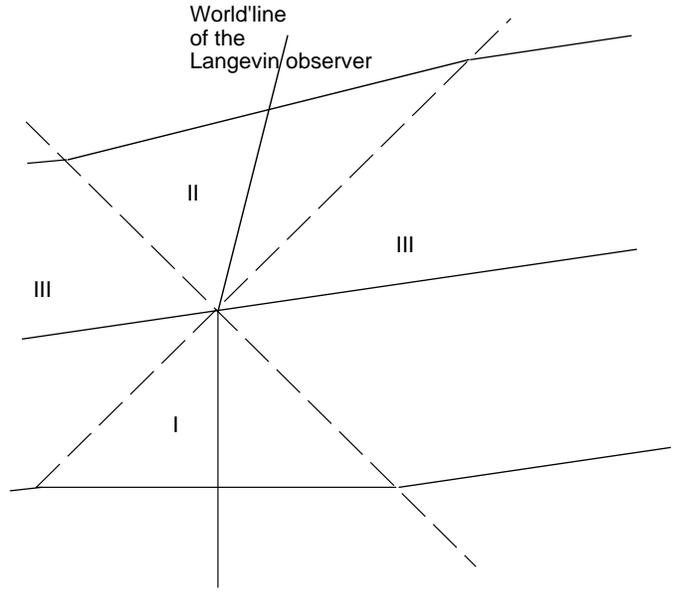}
\caption{World's line and (cuts of) space at various moments for the Langevin 
observer.
%(see Fig.6 for a perspective view). 
Its line cone is indicated by the  dashed lines.}
\label{Fig5}
   \end{figure}

It is well known that the solution of the celebrated " Langevin's twin 
paradox~" lies in geometry. I define a {\sl Langevin observer} as an 
inertial observer which is initially inertial, then suffers an 
instantaneous acceleration, and then is inertial again (Fig.\ref{Fig5}). Such an 
observer is able to meet his twin, which remained always inertial, 
with a different lapse of proper time. Is is often quoted (see, \eg, 
Misner et al.,    1973) %\cite{Misner}) 
that it is 
impossible to  define space globally for such an observer. Here I 
show that the synchronicity prescription applies perfectly and 
provides an unambiguous definition of space for this observer.
Thus I define the \spt ~trajectory of this  observer  as
$$x^{0}=\tau,x^{1}=x^{2}=x^{3}=0, \mbox{~~~~~for~~~~~}t<0,$$
$$x^{0}=c~\tau,x^{1}=s~\tau, x^{2}=x^{3}=0, \mbox{~~~~~for~~~~~}t>0,$$
with $c:=\cosh \psi$ and $s:=\sinh \psi$.

%%%%%%%%%%%%%%%%%%%%
%%%%%%%%%%%%%%%%%%%%
%%%%%%%%%%%%%%%%%%%%
%%%%%%%%%%%%%%%%%%%%
%%%%%%%%%%%%%%%%%%%%
%%%%%%%%%%%%%%%%%%%%

The light cone of the observer at $t=0$, $\calL^{0}$, 
divides the \spt ~into three 
parts I,~II and III (see  Fig.\ref{Fig5}) corresponding to the past, future 
and spatially related regions to the acceleration point. Studying the 
light-rays  from [to] the point  $x$ to [from] 
the observer gives the functions:\begin{itemize}
\item {\bf Region $I$}
%at proper times $\tau ^{+} $ and $\tau ^{-} $ defined by 
\begin{equation}
 \calN ^{\varepsilon} (x)=x^{0}+\varepsilon \sqrt{(x^{1})^{2}
 +(x^{2})^{2}+(x^{3})^{2}},
\end{equation}  
\begin{equation}
T(x)=t, R (x) =   \sqrt{(x^{1})^{2}+(x^{2})^{2}+(x^{3})^{2}}.\end{equation}  

\item {\bf Region $II$}
\begin{equation}
\begin{array}{l}  
\calN ^{\varepsilon}(x) = c~x^{0}-s~x^{1} \\   +\frac{\varepsilon}{c}  \sqrt{( 
c~x^{0}-s~x^{1})^{2} 
-(x^{0})^{2}+(x^{1})^{2}+(x^{2})^{2}+(x^{3})^{2}}  \\  
 = c~x^{0}-s~x^{1} +\frac{\varepsilon}{c} \sqrt{( 
s~x^{0}-c~x^{1})^{2}  +(x^{2})^{2}+(x^{3})^{2}} ,\end{array}
\end{equation}
$$ \begin{array}{l} T  (x)= c~x^{0}-s~x^{1},\end{array}$$
\begin{equation}
 \begin{array}{l} R(x)=\\    \sqrt{( c~x^{0}-s~x^{1})^{2} 
-(x^{0})^{2}+(x^{1})^{2}+(x^{2})^{2}+(x^{3})^{2}} \\  
 = \sqrt{( s~x^{0}-c~x^{1})^{2} + (x^{2})^{2}+(x^{3})^{2}}.\end{array} 
\end{equation}

\item {\bf Region $III$}
\begin{equation}
 \begin{array}{l} \calN ^{+}(x)= c~x^{0}-s~x^{1} \\  +   \sqrt{( 
s~x^{0}-c~x^{1})^{2}  +(x^{2})^{2}+(x^{3})^{2}},
\end{array}
\end{equation}
$$ \begin{array}{l}\calN ^{-}(x)=x^{0}  -\sqrt{(x^{1})^{2}+(x^{2})^{2}+(x^{3})^{2}},\end{array}$$
\begin{equation}
 \begin{array}{l}   2T (x)  =x^{0}+c~x^{0}-s~x^{1}\\ 
+\sqrt{(s~x^{0}-c~x^{1})^{2}  +(x^{2})^{2}+(x^{3})^{2}} \\
  -\sqrt{(x^{1})^{2}+(x^{2})^{2}+(x^{3})^{2}}  \end{array}
\end{equation} 
 and \begin{equation}
 \begin{array}{l} 2R (x)   =-x^{0}+c~x^{0}-s~x^{1} \\  
+\sqrt{(s~x^{0}-c~x^{1})^{2} + (x^{2})^{2}+(x^{3})^{2} } \\
 +\sqrt{(x^{1})^{2}+(x^{2})^{2}+(x^{3})^{2}}. \end{array} 
 \end{equation}
\end{itemize}

The surface $\Sigma _{\tau}$ of equation $T(x)=\tau$ defines space for 
the observer at proper time $\tau$.
%according to (Fig.\ref{Fig5b}). 
Let us 
consider its projection in the $(x^{0},x^{1})$ plane (Fig.\ref{Fig5}): 
\begin{itemize}
\item {\bf Region $I$}
\begin{equation}
\mbox   x^{0}=\tau.\end{equation}  
This is a straight horizontal line, where $ R (x) = x^{1} $.
\item {\bf Region $II$}
\begin{equation} c~x^{0}-s~x^{1}=\tau \end{equation}
or \begin{equation}
\tau/\cosh \psi=  x^{0}-\tanh \psi~x^{1}.\end{equation}
This is a line  inclined of $\psi$ with respect to the 
vertical, and thus orthogonal to the world line of the 
observer in that region. In this line, $$ R(x)= \sqrt{( c~x^{0}-s~x^{1})^{2} 
-(x^{0})^{2}+(x^{1})^{2} }  
= c~x^{1}-s~x^{0}.$$ 

\item {\bf Region $III$}
\begin{equation}
 \begin{array}{l} 2\tau =x^{0}+c~x^{0}-s~x^{1} \\ 
+\sqrt{(c~x^{0}-s~x^{1})^{2} -(x^{0})^{2}+(x^{1})^{2} }
  -  x^{1} \\
 = x^{0}+c~x^{0}-s~x^{1} +c~x^{1}-s~x^{0}   -  x^{1}, \end{array}
\end{equation}
 or \begin{equation}
\frac{\tau~\exp(\psi/2)}{\cosh(\psi/2)}   =- x^{0}+ x^{1}~\tanh 
(\psi/2). \end{equation} 
This is a line inclined of $\psi/2$ with respect to the vertical, 
\ie, at equal hyperbolic angle $\psi/2$ of  the two  previous 
lines (Fig.\ref{Fig5}). In this line, 
\begin{equation}
 \begin{array}{l} 2R (x)   =-x^{0}+c~x^{0}-s~x^{1} \\
+\sqrt{(c~x^{0}-s~x^{1})^{2} 
-(x^{0})^{2}+(x^{1})^{2} } +  x^{1} \\
= - x^{0}+c~x^{0}-s~x^{1} +c~x^{1}-s~x^{0}   + x^{1}.
 \end{array}
\end{equation} \end{itemize}

For the observer at an arbitrary moment, space is made of a plane 
circle  $S^{I}$ [or $S^{II}$] up to the light cone $\calL^{0}$ and is 
continued by a composite surface  $S^{III}$ beyond.
% as indicated in   (Fig.\ref{Fig5b}). 
Except at the single moment when the observer  experiences the 
instantaneous acceleration, space is not flat, nor homogeneous.

This is the simplest example where our prescription differs from the 
other one. As it is well known, it is impossible to extend the Fermi 
coordinates outside the conical regions. And no homogeneous hypersurfaces 
would be convenient. Thus, in this simple case,  our prescription is 
the only one   providing a reference frame associated 
to the observer valid in the whole \spt, to extend  the validity of 
his proper time, and to consider  unambiguous  synchronicity 
procedures (a similar conclusion has been reached by   Dolby and  
Gully, 2001). 
  
%%%%%%%%%%%%%%%%%%%%
%%%%%%%%%%%%%%%%%%%%
%%%%%%%%%%%%%%%%%%%%
%\begin{figure} 
%\psfig{file=Fig5b.eps,width=8.8cm}
%\caption{A cut of space (in perspective) for the Langevin observer at 
%(proper) time $\tau$. Fig.5 represents a cut of this figure.}
%\label{Fig5b}  \end{figure}
%%%%%%%%%%%%%%%%%%%%
%%%%%%%%%%%%%%%%%%%%
%%%%%%%%%%%%%%%%%%%%

\section{The Rindler  observer}\label{Rindler}
%%%%%%%%%%%%%%%%%%%%%%%%%%%%%

The  Rindler  observer in \mink ~has constant  acceleration $a$. Its velocity is defined 
by \begin{equation} 
u^{0}=\cosh (a ~\tau),~u^{1}=\sinh (a ~\tau) = ,~u^{2}= u^{3}= 0,
\end{equation} with  
acceleration $a^{0}=a~ \sinh (a~\tau) ,~a^{1}=a~ \cosh(a~ \tau) ,~a^{2}= a^{3}= 
0$ (with no loss of generality, I have taken the $x^{2}$ direction 
parallel to the acceleration). The world line has the  equation 
$x^{0}=a^{-1}~ \sinh( a \tau) ,~x^{1}=a^{-1}~\cosh( a \tau) ,~x^{2}= 
x^{3}= 0$, an hyperbola in \spt.
A light-ray from (to)  $x$ reach the observer at proper times 
$\tau   $   such that \begin{equation}\label{lray1}
[x^{0}-a^{-1}~\sinh (a \tau  )]^{2} 
=  [x^{1} -  a^{-1}~ \cosh ( a \tau   )]^{2} +(x^{2}) ^{2}+(x^{3}) ^{2}.
\end{equation} 
Since (anticipating) the solution requires  $ (x^{1})^{2}-(x^{0})^{2}>0$, 
we can introduce  the Rindler coordinates 
$$   x^{0}   :=a^{-1}~\exp(a~ \xi) ~\sinh (a~\eta) $$
 and \begin{equation}
x^{1}   :=a^{-1}~\exp( a~\xi) ~\cosh (a~\eta).  
 \end{equation}

\subsection{The problem in two dimensions}
%%%%%%%%%%%%%%%%%%%%%%%%%%%%%
 
I present first the problem in two dimensions, as it  is usually 
treated, since it is particularly simple and pedagogic. I further 
treat the complete 4-dimensional problem.

The condition (\ref {lray1}) becomes $$\cosh (a~\tau -a~\eta)=\cosh (a~ 
\xi),$$   or $ \tau -\eta =\pm\xi$. It results that
$$\calN  _{\varepsilon}=\eta +\varepsilon \norm{\xi},$$
$$T=\eta, R  =\norm{\xi}$$ and $N^{2}=a^{-2}~\exp (2a~\xi)= [x^{2}-t^{2}]$, 
so  that $N~dN=  x~dx- t~dt =2a^{-1}~ \exp (2a~\xi)~d\xi$.

The hypersurface (line)  $\Sigma _{\tau}$ has the equation $\eta =\tau$, which 
implies  $x^{0}=\tanh \tau ~x^{1}$,  a straight line through 
the origin. The   surfaces of constant PT-interval $\delta$ are 
the hyperbolae  of equation   $\xi =\delta$, or $(x^{1})^{2} -(x^{0})^{2}=a^{-2}~\exp 
(2a~\delta)$.

The forms \begin{equation}
dT=d\eta= \frac{x~dt-t~dx}{x^{2}-t^{2}} 
\end{equation} and \begin{equation}
dR=d\xi= \frac{-t~dt+x~dx}{x^{2}-t^{2}},
\end{equation}
so that  $n^{\varepsilon} = \frac{\varepsilon ~ dt+dx}{x+\varepsilon ~t}$.
%and  $$n^{+} =\frac{1}{x-t}~( -dt+dx).$$

\subsection{The four-dimensional  problem}
%%%%%%%%%%%%%%%%%%%%%%%%%%%%%

%Let us similarly write
%$ x^{0}   :=\exp( \xi) ~\sinh \eta$ and $ x^{1}   :=\exp( \xi) ~\cosh \eta$.
The solutions of (\ref {lray1})  are  given by 
\begin{equation}
% \begin{array}{l}
\cosh (a~\eta -a~\tau)\end{equation}
$$=\cosh (a~ \xi )+a^{2}~\exp (-a~\xi)~ 
[(x^{2}) ^{2}+(x^{3}) ^{2}] /2  :=\cosh  a~\xi ',
%\end{array} 
$$ where 
$$\exp (a~\xi '):=\cosh(a~ \xi) +\exp (-a~\xi)~a^{2}~ 
[(x^{2}) ^{2}+(x^{3}) ^{2}] /2$$
$$+\sqrt{(\cosh(a~  \xi) +\exp (-a~\xi)~~a^{2}~ 
[(x^{2}) ^{2}+(x^{3}) ^{2}]/2 )^{2}-1}>1.$$
This implies $\eta =\tau \pm \xi '$, and thus,  
$$\calN  _{\varepsilon}=\eta +\varepsilon ~  \xi '  ,$$
%$$\calN  _{-}=\eta -\mid \xi ' \mid ,$$ 
$$T= \eta,$$ 
$$R  =  \xi '   .$$ The forms \begin{equation}
dT=d\eta= \frac{x^{1} ~dx^{0}-x^{0}~d x^{1}}{(x^{1})^{2}-(x^{0})^{2}} 
\end{equation} and \begin{equation} 
dR=d\xi'=d\xi= \frac{-x^{0}~dx^{0}+ x^{1}~d x^{1}}{( 
x^{1})^{2}-( x^{0})^{2}},
\end{equation} so that  
$n^{\varepsilon} = \frac{\varepsilon ~ dx^{0} +dx^{1} }{x^{1} +\varepsilon 
~x^{0} }$ and  $N^{2}=a^{-2}~\exp (2a~\xi)=[(x^{1})^{2}-(x^{0})^{2}].$

The hypersurface  $\Sigma _{\tau}$ is the [flat] hyperplane  through 
the origin of  equation  $x^{0}=\tanh \tau ~x^{1}$ (it has as  
projection the   line seen in the previous section).
The (spatial) metric on $\Sigma _{\tau}$ is given by 
\begin{equation}
 \begin{array}{l} d\sigma^{2}=N^{2}~dR  
^{2}+(dx^{2})^{2}+(dx^{3})^{2}\\ =a^{-2}~
[d\exp(a~\xi)]^{2}+(dx^{2})^{2}+(dx^{3})^{2} \\
 =(dx^{1}/\cosh \eta)^{2}+(dx^{2})^{2}+(dx^{3})^{2},\end{array} 
\end{equation}
the latter form showing that its hypersurface is flat and homogeneous.

The   surfaces of constant PT-interval $\delta$ are 
given by  $ \xi'   =\delta$, or \begin{equation}
 \begin{array}{l}
2~a^{-1}~\cosh (a~\delta) ~\sqrt{(x^{1})^{2}-(x^{0})^{2}}-a^{-2}~\\= 
 (x^{1})^{2}-(x^{0})^{2} +(x^{2})^{2}+(x^{3})^{2}.
\end{array}
\end{equation} 

%\subsection{The Unruh effect}
%%%%%%%%%%%%%%%%%%%%%%%%%%%%%

This calculation  shows that the   widely used  Rindler coordinates
correspond in fact to the definition of   space and time introduced 
here for    the accelerated  observer, which    justifies their use. 
This sheds some light on  the Unruh  effect, which appears as a 
consequence of the  
different space-time splittings for the two observers (inertial and  
Rindler): they  associate different frequencies to the   same   state (\eg, the 
Minkowski inertial  vacuum). This has led   
Pauri  and Vallisneri (1999) %(\cite{Pauri})
 to invoque a   
classical (not quantum) origin for this effect, to be discussed further. 
  
\section{The cosmological observer}
%%%%%%%%%%%%%%%%%%%%%%%%%%%%%

Turning to cosmology, I consider  the \frl ~models, \ie, \spt s with spatial 
sections of  maximal symmetry. There exists a  special  system 
of coordinates in which   the  metric takes the form \begin{equation}
ds^{2}=A(\eta) ^{2}~(d\eta -[d\sigma ^{2}
-S(\sigma)^{2}~(d\alpha ^{2}+\sin ^{2}\alpha ~d\beta ^{2} ) ]),
\end{equation} where $A$ is the usual scale factor, 
    the expression between quotes is  the metric of 
a spatial section  with maximal symmetry (thus $\Ret$, $S ^{3}$ or $H^{3}$) 
and   $\eta$ is the 
{\sl conformal time}. Although different  systems of coordinates would be as 
well convenient,  I will perform  calculations with the coordinates 
$(\eta,\sigma,\alpha,\beta)$. 
%As it will appear, care must be taken with the use of these coordinates 
%since they are sometimes misleading.

I will   consider only  a   cosmological {\sl inertial} 
observer (CIO)  $O_{I}$ which  follows the line $\sigma=0$. The 
problem thus preserves spherical symmetry. All light-rays considered 
will be radial, and $\alpha,\beta$ will play no role. His proper time 
$\tau$  is  defined by $d\tau = A~d\eta$. The functions 
$\eta (\tau)$ and its inverse   $f$ such 
that $f[\eta (\tau)]:=\tau $ will play an important role.
Since  $\eta >0$, the CIO has a particle horizon and  $\calM_{0}$ is defined 
inside  it, \ie, by $\sigma <\eta$.

For the CIO,  \begin{equation}
\calN _{\varepsilon}( \eta ,\sigma)=f[\eta + \varepsilon ~ \sigma ],
\end{equation}   \begin{equation}\label{Tspt}
2T( \eta ,\sigma)=f[\eta + \sigma ]+ f[\eta - \sigma ],
\end{equation}  
\begin{equation}
2R ( \eta ,\sigma)=f[\eta + \sigma ]  -f[\eta - \sigma ].
\end{equation}  

Differentiation gives \begin{equation}
n_{\varepsilon}=A^{ \varepsilon}  
~ (d\eta +\varepsilon ~d \sigma),\end{equation}  
where I have defined $A^{\varepsilon}(\eta ,\sigma ):=A (\eta 
+ \varepsilon ~ \sigma )$. 
Sum and difference lead to \begin{equation}
dT=(A^{+}+A^{-})/2~d\eta +(A^{+}-A^{-})/2~d  \sigma
\end{equation} and  \begin{equation}
dR=(A^{+}-A^{-})/2~d\eta +(A^{+}+A^{-})/2~d \sigma,
\end{equation} and thus \begin{equation}
N^{2}(\eta ,\sigma )=\frac{A (\eta)^{2}}{A^{+}~A^{-}}.\end{equation}

The parallel transport of the velocity of the CIO along 
the light rays leads to \begin{equation}
\tilde{U}^{\varepsilon}=
\frac{1}{A^{\varepsilon}  }~[((A^{\varepsilon})^{2}+A^{2})~d\eta
+\varepsilon ~((A^{\varepsilon})^{2}-A^{2})~d \sigma ].
\end{equation}

\subsection{Space for the inertial cosmological observer}
%%%%%%%%%%%%%%%%%%%%%%%%%%%%%

Space, \ie, the surface  $\Sigma _{\tau}$, has the equation 
\begin{equation} f[\eta   +\sigma]+f[\eta  -\sigma]=2\tau,~\mbox{with~} 
\sigma <\eta, \end{equation} which is {\sl not} the 
surface $\eta=C^{te}$ (except in the case with no  expansion, where $\eta=\tau$).
In other worlds, space for the  CIO   is not  a  spatial section with maximal 
symmetry:  the cosmic  expansion breaks the spatial 
homogeneity  (although not its isotropy when the observer is 
inertial). The spatial 
sections $\eta =C^{te}$, sometimes quoted as ` space ` do not verify the 
 synchronicity condition (they verify a kind of  
 synchronicity condition, but in the conformal time which has no 
physical relevance for the observer,  rather than  in its proper time). 

Also, the cosmic expansion imprints a curvature onto space: even when  
\spt ~admits   spatial  sections  of constant curvature (like for 
instance flat in the \EdS ~case), this is not the case for the  
space. Space is limited by the horizon $\sigma =\eta$, or $T=R$. On the 
horizon, $A^{-}\rightarrow 0$ and thus $N\rightarrow \infty$. Space  
tends to become null (light-like).

The PT-interval to a point of this surface is 
\begin{equation}R =f[\eta + \sigma ]  /2-f[\eta - \sigma ]/2=f[\eta 
+ \sigma ]-T.\end{equation}  

{\bf  The \EdS ~model}

\begin{figure} 
\psfig{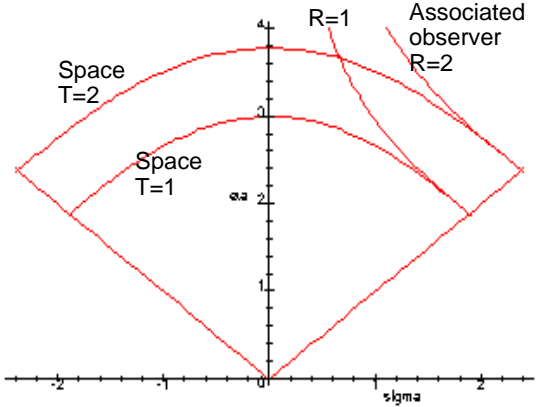}
\caption{Space ($T=Ct$) and world line of associated observers 
($R=Ct$) for the CIO in \EdS ~\spt.}
\label{Fig6}   \end{figure}

To  check this  in more detail, I consider  the example of {\sl 
Einstein-\dS ~\spt}, with 
$A(\tau)=\tau^{2/3}=(\eta /3)^{2} $ and flat spatial sections. 
In this case, $\eta (\tau)=3~\tau^{1/3}$ and $f(y)=(y/3)^{3}$. Thus, 
 $\Sigma _{\tau}$ has the equation (Fig.\ref{Fig6})   \begin{equation}
27~\tau   =\eta ~(\eta  ^{2}+3~\sigma ^{2}),\end{equation} and not 
$\tau =\eta ^{3}/27$, the  equation of a flat spatial section. We 
have $N(\eta,\sigma)=\frac{\eta ^{2} }{\eta ^{2} -\sigma ^{2} }=[1+ 
^{3}\sqrt{\frac{\tau +R}{\tau -R}}+ 
^{3}\sqrt{\frac{\tau -R}{\tau +R}}]/4$.
The PT-interval to a point $x$ is  $R =\frac{\sigma}{27}~(3~\eta ^{2}+\sigma ^{2})$, 
different from the usual proper distance at time $\eta$, namely 
$\sigma ~A(\eta) =\sigma \eta ^{2}/9$.
 Space, at the position of the observer, is orthogonal to its world 
line, as expected. 
%More interesting is the behaviour of space far 
%from the CIO: taking the limit $R \rightarrow \infty$ or, equivalently,  
%  $\sigma \rightarrow \infty$, we obtain 
%  that the equation of $\sigma _{\tau}$ becomes equivalent to

%%%%%%%%%%%%%%%%%%%%
%%%%%%%%%%%%%%%%%%%%
%%%%%%%%%%%%%%%%%%%%
%%%%%%%%%%%%%%%%%%%%
%%%%%%%%%%%%%%%%%%%%
%%%%%%%%%%%%%%%%%%%%

{\bf Associated observers}
 
As we have seen, any observer has a class of associated observers. 
We point our that these observers are not the comoving observers 
defined by $\sigma=C^{te}$, which  obey the equation $
dT=\frac{A^{+}+A^{-}}{A^{+}-A^{-}}~dR$. On the other hand, the associated 
observers keep a  constant PT-interval and obey the equation \begin{equation}
dR=(A^{+}-A^{-})/2~d\eta +(A^{+}+A^{-})/2~d\norm{\sigma}=0. 
\end{equation}
%Their proper time is given by 
%dt^{2} =N^{2}~dT^{2}=N^{2}~[
%dT=(A^{+}+A^{-})/2~d\eta +(A^{+}-A^{-})/2~d\norm{\sigma}
%] =
 
An associated observer  at the horizon is seen by the CIO with a 
 redshift $z^{+}\rightarrow \infty$.
 
{\bf Time and distances}

All measurements made by an    observer, local or not, refer to his   proper 
time. Thus, when a  CIO considers an event in \spt, the relevant time to measure 
durations, or to date the event, is not $t$ or $\eta  $ but $T$ defined 
above (I recall that $T$ and $t$ coincide {\sl on the world line of 
the CIO}). 

On the other hand, the proper distance is intended to measure the interval 
between two  objects considered simultaneously, \ie, at a common value 
of time. But, again,  no observer has access to  the conformal time 
$\eta$. Thus,  simultaneity (not    absolute, but only 
relative to the observer) must be defined not by $\eta$ but by $T$ as we have 
explained. This leads to use  the {\sl proper time  distance} (that I call 
differently  to avoid  confusion) between two objects, calculated by 
integration of the  metric element,   not along a spatial 
section $t=C^{te}$ (or $\eta = C^{te}$), but along $\Sigma _{T}   $, \ie, 
\begin{equation} 
 d_{PT}(g)=\int _{\Sigma _{T} }~ds=\int _{\Sigma _{T} } N~dR.
\end{equation}   The  PT-distance
is thus   really the distance between two objects in  
{\sl space}, at a given moment for the CIO.

%Let us, for instance, consider a comoving galaxy $g$ in the usual sense, 
%\ie, with $d\sigma =0$. Usually, its  proper distance (to the CIO) 
%at time $t$ is  calculated as $d_{proper}(g)=\int _{t=Ct}~ds$.
%Its PT-distance is however given by 

Thus, for the CIO, $T$ and $R$ appear as convenient coordinates 
to measure space and time.
 
\subsection{Inertial   observer in \dS ~\spt}
%%%%%%%%%%%%%%%%%%%%%%%%%%%%%

The  case of    the  \dS ~\spt ~is particularly interesting since, 
because of its maximal symmetry it has been widely considered as a 
frame  for  quantization.
As it is well known, it admits different slicings, with  spatial sections of 
constant curvature  with positive, zero or negative signs. To each 
slicing corresponds a convenient system of coordinates. Here I chose 
the slicing where spatial sections have constant positive curvature 
since it covers the whole \spt ~(\dS ~hyperboloid). 

The metric is written as  \begin{equation}
\begin{array}{l}
ds^{2}=dt^{2}-\rho ^{2}~(\cosh \rho ^{-1}  t)^{2}~[d\sigma ^{2}
+\sin ^{2} \sigma ~d\Omega ^{2}]\\
=A^{2}(\eta)~[d\eta ^{2}- d\sigma ^{2} -\sin ^{2} \sigma ~d\Omega 
^{2}], \end{array}
\end{equation}
where the conformal time $\eta =\rho \tan ^{-1}[\sinh(\rho 
^{-1}~t)]$ and $A(\eta)=\rho/\cos(\eta /\rho)$ ($\rho$ is a constant 
characterizing the constant negative curvature of \spt).

The chosen CIO has proper time $t=\rho ~\sinh ^{-1}[\tan (\eta/\rho)]$ 
so that $f(y):=\rho ~\sinh ^{-1}[\tan (y/\rho)]$.
Thus, \begin{equation}
\calN _{\varepsilon}(t,\sigma)=\rho~\ln 
\frac{1+S^{\varepsilon} }{C^{\varepsilon}},\end{equation}
where $S^{\varepsilon}:=\sin \frac{\eta +\varepsilon 
\norm{\sigma}}{\rho}$, $C^{\varepsilon}:=\cos \frac{\eta +\varepsilon 
\norm{\sigma}}{\rho}$. We obtain\begin{equation}
2T(\eta,\sigma)=\rho~\ln \frac{\cos (\sigma /\rho) +\sin ( 
\eta/\rho)}{\cos ( \sigma /\rho) -\sin ( \eta/\rho)},
\end{equation}
\begin{equation}
2R(\eta,\sigma)=\rho~\ln \frac{\cos (\eta/\rho) 
+\sin ( \sigma/\rho)}{\cos( \eta /\rho) -\sin (\sigma /\rho)}. \end{equation}
    
Space for the CIO at (proper) time $\tau$ is given by     
\begin{equation}
\cos (\sigma /\rho) +\sin (  \eta/\rho)=\exp(2\tau/\rho)~[\cos (\sigma /\rho) 
-\sin (  \eta/\rho)]],   
\end{equation}
or \begin{equation}
\cos (\sigma /\rho) ~\tanh(\tau/\rho)=\sin (\rho ^{-1}~t),        \end{equation}
which, again, is {\sl not} the surface $t=C^{te}$ of constant (positive) curvature.
Thus we claim that quantization  must be performed with this surface 
and its orthogonal time. This will be studied in a forthcoming paper.
 
 \section{Discussion}

The  prescription  based on synchronicity
  defines space without ambiguity for  any given observer, inertial or not,
in arbitrary \spt ~(without multi-crossing of null geodesics),  
including \mink ~and the \frl ~models. Space is relative 
to the observer (it is in general  different for an other observer), and   well 
defined at each instant of   its world line. 
This provides  a   foliation of \spt, valid for this observer, which 
may be interpreted as   a class of   {\sl  canonically associated}   
observers or a " kinematics " of \spt ~(Smarr and York 1978).  
This provides also a natural reference frame, \ie, global    space and time 
coordinates in the whole \spt, which remains 
Minkowskian along the world line of the observer (thus, time   coincides 
with its proper time there)   and which is pertinent for physical measurements. 
In many  cases 
(in particular  for Rindler observers; see all references 
concerning the Rindler effect, and Sriramkumar   and Padmanabhan, 
1999),   
% \cite{paths}), 
the coordinate system 
introduced here coincides with that used in the 
literature   with no other justification than   being    "~natural~", 
and thus  
provides an {\sl a posteriori} justification. Also,  the prescription presented here 
appplies to a range wider than other reference frames.

Application to \mink ~confirms that space and time differ for 
inertial observers with different velocities. It provides an 
unambiguous and global definition of space and time for the Langevin observer, 
for which the other prescriptions do not apply. 
Applied to the Rindler observer (with uniform acceleration), it 
leads to  space and time coordinates which  coincide with the usual 
Rindler coordinates, thus providing a justification of their use. The 
corresponding   interpretation of the Unruh effect   
involves  the observer-dependant 
character  of   space and time.

In cosmology,  this prescription provides,  for  the inertial observer in the 
general \frl ~model,  an unambiguous definition of 
space, which {\sl  does not coincide with a spatial section of maximal 
symmetry}. Thus, in the \frl ~models, no inertial observers "~sees~" a  
homogeneous  space. The lack of homogeneity of space is due to the    curvature
corresponding to  the    expansion law. In particular, 
space  is not flat  nor homogeneous (although 
the inertial character of the observer preserves its isotropy)  in the 
\EdS ~model, sometimes  called a  "~flat universe~"~!
I have also calculated space for the inertial observer in \dS ~\spt, 
which, again, is not a  hypersurface of maximal symmetry.

These results    do  not modify the cosmological formulae when they  are 
expressed in  a covariant form and do not  involve  a definition of space. 
However they change those interpretations of   observational  
results, which involve a reference to 
space (like " {\sl space} is homogeneous, flat " etc.). This modifies also the 
interpretation of the usual  {\sl proper distance}: it  does not appear as the 
proper spatial interval between two  events occurring  at the same time, but 
rather as a mixed interval between two  events which are not synchronous for 
the observer which  performs  the  measurement (they would be synchronous if 
the observer were using a watch  indicating " conformal time " rather 
than " proper time "). Thus I introduce a "~proper time-distance~", 
which  represents a  spatial interval  between two events which are 
synchronous  for the  observer. It corresponds to the result of a practical
measurement  that the observer may perform with his watch indicating  
his proper time. Its  value differs,  in general, form the usual proper distance.  

This  prescription for space could  have important  implications for 
interpreting 
quantum effects in curved \spt, and/or from the point of  view of non 
inertial  observers. Its application to the Rindler observer 
confirms the usual results of the  Unruh effect, and allows a clearer 
comprehension. In other cases,  the prescription adopted here
differs from  most attempts 
   up to  now, since the use of  spatial sections with 
maximal symmetry   (rather than space as defined 
here) does not obey the synchronicity requirements. This suggest a new 
examination of   quantum effects  in curved \spt, to be 
discussed in a forthcoming paper. Also, subsequent work will explore 
in more detail  the application of this procedure to  observers with arbitrary 
acceleration   and rotation.

 \end{document}